\documentclass[fleqn,usenatbib]{mnras}

\usepackage{newtxtext,newtxmath}

\usepackage[T1]{fontenc}

\DeclareRobustCommand{\VAN}[3]{#2}
\let\VANthebibliography\thebibliography
\def\thebibliography{\DeclareRobustCommand{\VAN}[3]{##3}\VANthebibliography}

\usepackage{graphicx}	
\usepackage{amsmath}	
\usepackage{enumitem}   
\usepackage{xcolor}
\usepackage{xspace}

\usepackage[switch,columnwise]{lineno}

\defcitealias{DES:2018crd}{C19}

\newcommand{\orcid}[1]{} 

\newcommand{\om}{\Omega_{\rm m}}

\newcommand{\lnas}{\ln(10^{10} A_s)}
\newcommand{\mwl}{{\rm M}_{\rm WL}}
\newcommand{\lob}{\lambda^{\rm ob}}
\newcommand{\ltr}{\lambda^{\rm tr}}
\newcommand{\zob}{z^{\rm ob}}
\newcommand{\ztr}{z^{\rm tr}}
\newcommand{\mmin}{M_{\rm min}}
\newcommand{\sintr}{\sigma_{\rm intr}}
\DeclareMathOperator{\logten}{log_{10}}
\newcommand{\diff}{\ensuremath{{\rm d}}}

\title[Cluster Cosmology: Abundance, Lensing \& Clustering]{Cosmological constraints from abundance, weak-lensing and clustering of galaxy clusters: application to the SDSS}

\author[A. Fumagalli et al.]{A.~Fumagalli$^{1,2,3,4}$\thanks{E-mail: alessandra.fumagalli@inaf.it}, M.~Costanzi$^{1,3,2}$,  A.~Saro\orcid{0000-0002-9288-862X}$^{1,2,3,5,6}$, T.~Castro\orcid{0000-0002-6292-3228}$^{2,3,5,6}$, S.~Borgani\orcid{0000-0001-6151-6439}$^{1,3,2,5,6}$ 
\\
$^{1}$ Dipartimento di Fisica - Sezione di Astronomia, Universit\'a di Trieste, Via Tiepolo 11, 34131 Trieste, Italy\\
$^{2}$ INAF-Osservatorio Astronomico di Trieste, Via G. B. Tiepolo 11, 34143 Trieste, Italy\\
$^{3}$ IFPU, Institute for Fundamental Physics of the Universe, via Beirut 2, 34151 Trieste, Italy\\
$^{4}$ University Observatory, Faculty of Physics, Ludwig-MaximiliansUniversität, Scheinerstr. 1, 81679 Munich, Germany\\
$^{5}$ INFN, Sezione di Trieste, Via Valerio 2, 34127 Trieste TS, Italy\\
$^{6}$ ICSC - Italian Research Center on High-Performance Computing, Big Data and Quantum Computing
}

\date{Accepted XXX. Received YYY; in original form ZZZ}

\pubyear{2023}

\begin{document}

\label{firstpage}
\pagerange{\pageref{firstpage}--\pageref{lastpage}}
\maketitle

\begin{abstract}
   The clustering of galaxy clusters is a powerful cosmological tool, which can help to break degeneracies between parameters when combined with other cosmological observables.  We aim to demonstrate its potential in constraining cosmological parameters and scaling relations when combined with cluster counts and weak lensing mass information, using as a case study the redMaPPer cluster catalog derived from the Sloan Digital Sky Survey (SDSS).
   We extend the analysis of number counts and weak lensing signal performed by Costanzi et al. 2019a, with the addition of the real-space 2-point correlation function. We derive cosmological and scaling relation posteriors for all the possible combinations of the three observables to assess their constraining power, parameter degeneracies, and possible internal tensions.
   We find no evidence for tensions between the three data set analyzed. We demonstrate that the inclusion of the cluster clustering statistic can greatly enhance the constraining power of the sample thanks to its capability of breaking the $\om$\,--\,$\sigma_8$ degeneracy characteristic of cluster abundance studies. In particular, for a flat $\Lambda$CDM model with massive neutrinos, we obtain $\om=0.28 \pm 0.03$ and $\sigma_8 = 0.82 \pm 0.05$, a 33\,\% and 50\,\% improvement compared to the posteriors derived combining cluster abundance and weak lensing analyses. Our results are consistent with cosmological posteriors from other cluster surveys, as well as with Planck CMB results and DES-Y3 galaxy clustering and weak-lensing analysis.
\end{abstract}




\section{Introduction} \label{sec:intro}
Clusters of galaxies are known to be powerful probes for studying the geometry and evolution of the large-scale structure of the Universe. Originating from high-density regions in the initial matter density field, they grow through a hierarchical process of accretion and merging of small objects into increasingly larger systems, and they currently represent the most massive virialized objects in the Universe \citep[for reviews, see e.g.][]{Allen:2011zs,Kravtsov:2012zs}. Over the past two decades the statistical tool mostly adopted to extract cosmological information from cluster catalogs is the study of their abundance in mass, namely the number counts \citep[e.g.,][]{Borgani:2001ir,Vikhlinin:2008ym,Planck:2013lkt,Bocquet:2015pva,DES:2020cbm}, which allows to derive constraints on the average matter density ($\om$) and the amplitude of density fluctuations ($\sigma_8$) in the Universe. Several examples can be provided about cosmological constraints from cluster counts analyses, both from optical cluster catalogs \citep[e.g.,][]{DES:2018crd,DES:2021wwk,Lesci:2020qpk}, X-ray catalogs \citep[e.g.,][]{Mantz:2009fw,Mantz:2014paa,Schellenberger:2017usb,Chiu:2022qgb} or millimeter catalogs obtained through the Sunyaev-Zeldovich signature \citep[e.g.,][]{Planck:2015lwi,SPT:2018njh}, as well as from the combination of multi-wavelength data \citep[e.g.,][]{DES:2020cbm} or with different cosmological probes, such as CMB \citep{Salvati:2021gkt}.

The study of the clustering properties of galaxy clusters provides another mean for their cosmological exploitation \citep{Borgani:1998sfa,Moscardini:1999ba,Estrada:2008em,Marulli:2018owk}. Such a probe offers several advantages: the fact that clusters are highly biased tracer of the matter density field, their detectable clustering signal on large scales where linear theory is applicable, the influence of cosmological parameters on the bias-mass relation \citep{Mo:1995cs,Tinker:2010my}, and the minimal impact of baryonic effects on clustering statistics \citep{Castro:2020yes} collectively contribute to the enhanced constraining power and reliability of cluster clustering as a valuable tool for cosmological investigations. Although the available statistics is still too low to use the clustering of clusters as a competitive stand-alone probe, it has been shown to be a valuable source of information for breaking parameter degeneracies when combined with other observables such as cluster counts \citep{Schuecker:2002ti,Majumdar:2003mw,Sereno:2014eea,Sartoris:2015aga}, and to improve the calibration of mass-observable relations \citep{Mana:2013qba,DES:2020uce,Lesci:2022owx}. 

While cluster abundance and halo bias can be predicted with good accuracy as a function of mass from first principles or simulations \citep{Mo:1995cs,Sheth:1999mn,Tinker:2010my,Euclid:2022dbc}, cluster masses are not directly observable. The mass inference process must in fact rely on observational proxies that exhibit correlations with mass, such as properties of galaxies (richness, velocity dispersion) or of the gas within the cluster (total gas mass, temperature, pressure). The scaling-relations linking these observables to cluster masses provide a statistical estimate of the latter, but require a careful calibration to avoid biases in the cosmological inference \citep{Kravtsov:2012zs,Pratt:2019cnf}.  Weak gravitational lensing measurements, based on the estimate of the tangential alignment of background galaxies around foreground clusters due to gravitational lensing, provide the gold-standard technique for estimating cluster masses \citep{SDSS:2007xgc,Hoekstra:2013via,vonderLinden:2014haa,Murata:2017zdo,Simet:2016mzg,DES:2016opl}. The advantages in using weak lensing mass information is that weak lensing measurements capture both dark and baryonic matter, and do not rely on assumptions about the cluster's dynamical state. However, the calibration process is challenging due to various biases and observational uncertainties. Systematic errors, including shear and photometric redshift biases, halo triaxiality, and projection effects, hamper the interpretation of the weak lensing measurements and dominate the overall error budget of the mass calibration \citep{vonderLinden:2014haa,Hoekstra:2015gda,Simet:2016mzg,Miyatake:2018lpb,DES:2022qkn}.

Cosmological constraints obtained from cluster counts and cluster clustering, as well as the accuracy of the mass calibration from weak lensing, are expected to largely improve with the most recent and forthcoming wide-surveys such as the Dark Energy Survey\footnote{\url{https://www.darkenergysurvey.org}} \citep{DES:2005dhi}, eROSITA\footnote{\url{ http://www.mpe.mpg.de/eROSITA}} \citep{predehl2014}, \textit{Euclid}\,\footnote{\url{http://sci.esa.int/euclid/}} \citep{EUCLID:2011zbd} or Vera C. Rubin Observatory LSST \footnote{\url{https://www.lsst.org/}} \citep{LSSTDarkEnergyScience:2012kar}. Nevertheless, the cosmological gain provided by the clustering  statistics has not been thoroughly investigated in currently available cluster catalogs. To this purpose, we derive cosmological constraints from the combination of abundances, weak lensing, and clustering of
galaxy clusters from the redMaPPer cluster catalog in the the Sloan Digital Sky Survey data release 8 \citep[SDSS DR8,][]{SDSS:2011gir}; the analysis of cluster abundances and weak lensing is based on the work made by \citet[][hereafter C19]{DES:2018crd}.
The aim is to show how the inclusion of clustering does help both to narrow uncertainties on the inference of cosmological parameters from cluster counts and weak lensing masses, and to reduce biases and systematic uncertainties in the mass calibration.

This paper is structured as follows. In Sect.~\ref{sec:theory} we present the theoretical framework to model the three statistics considered here, i.e., cluster counts, cluster clustering, and weak lensing masses, as well as the likelihood models adopted to infer cosmological and mass-observable relation parameters. In Sect.~\ref{sec:data} we describe the data analyzed in this work. In Sect.~\ref{sec:results} we present the results and in Sect.~\ref{sec:conclusions} we discuss our conclusions.

\section{Theory and methods} \label{sec:theory}
In this section, we introduce the theoretical formalism to describe number counts, weak lensing masses, and cluster clustering, along with their covariance matrices. We also describe the likelihood function adopted for the parameter inference.

\subsection{Cluster number counts}
We model the cluster number counts in the $i$-th richness bin and $j$-th redshift bin as
\begin{equation}
\label{eq:numbercounts_obs}
    \begin{split}
     \langle N(\Delta \lob_i, \Delta \zob_j) \rangle = & \int_0^\infty \diff \ztr \;\Omega_{\rm mask}(\ztr)\,\frac{\diff V}{\diff z\, \diff \Omega}(\ztr) \,  \\
     & \times \langle n(\ztr, \Delta \lob_i) \rangle \int_{\Delta \zob_j} \diff \zob P(\zob\,|\,\ztr, \Delta \lob_i)\,,        
    \end{split}  
\end{equation} 
where $\Omega_{\rm mask}$ is the redshift-dependent survey area (see \citetalias{DES:2018crd}), $\diff V / \diff \Omega \, \diff z$ is the comoving volume element per unit redshift and solid angle. Also, $P(\zob\,|\,\ztr, \Delta \lob_i)$ is the probability distribution of assigning an observed redshift $\zob$ given the true redshift $\ztr$ of a cluster and observed richness, to account for the observational scatter due to the photometric uncertainty; as described in \citetalias{DES:2018crd}, it is modeled as a Gaussian distribution with mean $\ztr$ and a redshift and richness-dependent variance. Lastly, the term $\langle n(\ztr, \Delta \lob_i) \rangle$ describes the expected number density of halos in the $i$-th richness bin, and is given by
\begin{equation}
\label{eq:massfunc_obs}
     \langle n(\ztr, \Delta \lob_i) \rangle = \int_0^\infty \diff M \; \frac{\diff n}{\diff M}(M,\ztr) \int_{\Delta \lob_i} \diff \lob P(\lob\,|\,M,\ztr)\,,  
\end{equation} 
where $\diff n/ \diff M$ is the halo mass function, and $P(\lob\,|\,M,\ztr)$ is the observed richness-mass relation,
\begin{equation}
\label{eq:mor}
     P(\lob\,|\,M,\ztr) = \int_0^\infty \diff \ltr \, P(\lob\,|\,\ltr,\ztr)\,P(\ltr\,|\,M,\ztr)\,.
\end{equation} 
The term $P(\ltr\,|\,M,\ztr)$ is the redshift-dependent intrinsic richness-mass relation; following \citetalias{DES:2018crd}, we modeled it as a skew-normal distribution with mean and variance given by
\begin{align}
    & \langle \ltr | M \rangle  = \left ( \frac{M - \mmin}{M_1-\mmin}\right )^\alpha\,, \label{eq:mor_mean}\\
    & \sigma^2_{\ln \ltr}  = \sintr^2 + \frac{\langle \ltr | M \rangle - 1}{\langle \ltr | M \rangle^2}\,. \label{eq:mor_var}
\end{align}
Here, $\mmin$ is the minimum mass for a halo to form a central galaxy,
while $M_1$ is the characteristic mass at which halos acquire one
satellite galaxy. The variance is composed by an intrinsic scatter, $\sintr$, plus a Poisson contribution.

The other term in Eq.~\eqref{eq:mor}, i.e., $P(\lob\,|\,\ltr,\ztr)$, represents the observational scatter in the richness-mass relation, due to photometric noise, uncertainties in background subtraction, and projection/percolation effects. Projection effects occur when multiple foreground and background objects along the same line of sight are mistakenly associated to a galaxy cluster, increasing the apparent richness of the cluster. As a consequence, percolation is the reduction in the count of member galaxies for clusters that are masked by these projection effects. Following \citetalias{DES:2018crd}, the background subtraction and photometric noise terms are modeled as Gaussian components, while projection and percolation effects as an exponential and uniform distribution, respectively. The corresponding model parameters have been calibrated by means of real and simulated data analysis (see \citetalias{DES:2018crd} for details). 

The covariance matrix associated to cluster number counts is computed analytically following the model by \citet{Hu:2002we}, validated in \citet{Euclid:2021api}. Such a model accounts for the effects of shot-noise and sample variance, plus a contribution due to uncertainty in the miscentring corrections, as described in \citetalias{DES:2018crd} (see their Appendix A).

\subsection{Weak lensing masses}
Following \citetalias{DES:2018crd}, for the mass calibration we rely on the mean cluster mass measurements from  the stacked shear analysis of \citet{Simet:2016mzg}, rather than modeling directly the cluster shear profiles. Similarly to number counts, the expectation value of the mean cluster mass within the $i$-th richness bin and $j$-th redshift bin is given by
\begin{equation}
    \label{eq:mean_mass_wl}
    {\overline M}(\Delta \lob_i, \Delta \zob_j) = \frac{\langle M^{\rm tot}(\Delta \lob_i, \Delta \zob_j)\rangle }{\langle N(\Delta \lob_i, \Delta \zob_j) \rangle}\,,
\end{equation}
where $\langle M^{\rm tot} \rangle$ is the total mass associated to clusters identified in given redshift and richness intervals, given by
\begin{equation}
\label{eq:mtot_obs}
    \begin{split}
     \langle M^{\rm tot}(\Delta \lob_i, \Delta \zob_j) \rangle = & \int_0^\infty \diff \ztr \;\Omega_{\rm mask}(\ztr)\,\frac{\diff V}{\diff z\, \diff \Omega}(\ztr) \, \\
     & \times  \langle M\,n(\ztr, \Delta \lob_i)\rangle \int_{\Delta \zob_j} \diff \zob P(\zob\,|\,\ztr, \Delta \lob_i)\,,    
    \end{split}  
\end{equation} 
with
\begin{equation}
\label{eq:M_massfunc_obs}
     \langle M\,n(\ztr, \Delta \lob_i) \rangle = \int_0^\infty \diff M \; M \frac{\diff n}{\diff M}(M,\ztr) \int_{\Delta \lob_i} \diff \lob P(\lob\,|\,M,\ztr)\,. 
\end{equation}

\subsection{Cluster clustering}
We model the real-space 2-point correlation function in the $a$-th radial bin and $j$-th redshift bin as \citep{Matarrese:1996bc}:
\begin{equation} \label{eq:2pcf_obs}
    \langle \xi(\Delta r_a, \Delta \zob_j)\rangle = \int \frac{\diff k\,k^2}{2\pi^2} \left \langle \overline{b}\,\sqrt{P_{\rm m}(k)} \right \rangle^2_j \, W_a(k)\,,
\end{equation}
where 
\begin{equation}
    \begin{split}
	 \left \langle \overline{b}\,\sqrt{P_{\rm m}(k)} \right \rangle_j & = \frac{1}{\langle N(\Delta \zob_j)\rangle} \int_0^\infty \diff \ztr \, \Omega_{\rm mask}(\ztr)\,\frac{\diff V}{\diff \Omega\,\diff z}(\ztr) \\
     & \times  n(\ztr) \,  \overline{b}(\ztr)  \, \sqrt{P_{\rm m}(k,\ztr)} \, \int_{\Delta \zob_j} \diff \zob P(\zob\,|\,\ztr, \Delta \lob_i)\,.    
    \end{split}
\end{equation}
In the above equation, $\overline{b}(\ztr)$ is the effective bias for all clusters having an observed richness above a given threshold value $\lob_{\rm th}$:
\begin{equation}
    \begin{split}
     \overline{b}(\ztr)  = \frac{1}{n(\ztr)} & \int_0^\infty \diff M \frac{\diff n}{\diff M}(M,\ztr)\,b(M,\ztr) \\ \times & \int_{\lob_{\rm th}}^\infty \diff \lob P(\lob\,|\,M,\ztr)\,.
    \end{split}
\end{equation}
Similarly, $N(\Delta \zob_j)$ and $n(\ztr)$ are given by Eqs.~\eqref{eq:numbercounts_obs} and~\eqref{eq:massfunc_obs}, with integrals over the richness bins substituted with integrals above the richness threshold \footnote{As explained in Sect.~\ref{sec:data}, we do not bin over richness to prevent the signal from being too low. }.

Finally, the term $W_a(k)$ in Eq.~\eqref{eq:2pcf_obs} is the spherical shell window function, given by
\begin{equation} \label{eq:shell_window}
     W_a(k) = \int \frac{\diff^3r}{V_a} j_0(kr) = \frac{r_{a,+}^3 W_{\rm th}(k r_{a,+}) - r_{a,-}^3 W_{\rm th}(k r_{a,-})}{r_{a,+}^3 - r_{a,-}^3}\,,
\end{equation}
where the $W_{\rm th}(kr)$  is the top-hat window function, $V_a$ is the volume of the $a$-th spherical shell, and $r_{a,-},\,r_{a,+}$ are the extremes of the separation bin. 

To take into account the uncertainties in the photometric redshifts, the matter power spectrum has to be modified as \citep{Sereno:2014eea}
\begin{equation}
    P_{\rm m}'(k) = P_{\rm m}(k)\,\frac{\sqrt{\pi}}{2\,k \sigma}\,{\rm erf}(k \sigma)\,,
\end{equation}
where $\sigma$ depends on the photo-$z$ error $\sigma_z$ as
\begin{equation}
    \sigma = \frac{c\,\sigma_z}{H(z)}\,.
\end{equation}
The effect of the photo-$z$ uncertainty is to decrease the correlation on small scales, while increasing it at large scales, around the BAO peak. This allows us to neglect the Infrared Resummation \citep[IR,][]{Senatore:2014via,Baldauf:2015xfa}, otherwise needed to correct for the broadening and shift of the peak due to non-linear damping. Moreover, the damping produced by photo-$z$ errors is similar to that produced by redshift-space distortions, and at this level of statistics it is the dominant source of error; therefore, real-space analysis should introduce minimal error compared with the more proper modeling in redshift-space \citep{Sereno:2014eea}.

Lastly, when measuring the 2-point correlation function from data, it is necessary to make assumptions about the cosmology to convert redshifts into distances. However, since it is not feasible to assume the true cosmology in the measurements, a parameter is introduced in the model (specifically, into Eq.~\ref{eq:shell_window}) to account for geometric distortions in the model \citep{Marulli:2012na,Marulli:2015jil}:
\begin{equation}
    \xi(r) \longrightarrow \xi(\alpha r)\,
\end{equation}
with
\begin{equation}
    \alpha = \frac{D_V}{r_s} \frac{r_s^{\rm fid}}{D_V^{\rm fid}}\,,
\end{equation}
where $r_s$ is the position of the sound horizon at decoupling, $D_V$ is the isotropic volume distance, and the label ``fid'' indicates the quantities evaluated at the fiducial cosmology, assumed in the measurement process.

We describe the clustering covariance matrix by applying the semi-analytical model presented and validated in \citet[][see equation 27]{Euclid:2022txd}. The model consists of a Gaussian covariance plus a low-order non-Gaussian term.  Since in this work we only analyze low-redshift clusters, the model works well without the need of fitting additional parameters, which are required to correct the covariance at high redshift. 

\subsection{Likelihood function}
We consider the three statistics as independent. Following the assumption of \citetalias{DES:2018crd}, we assume cluster counts and weak lensing data to be uncorrelated, given that the dominant systematics in the latter (shape noise, multiplicative shear and photo-$z$ bias) do not affect number counts. For the same reason we assume weak lensing to not correlate with clustering either.  Moreover, the scales considered for the two observables in this work do not overlap. Finally, we show in Appendix~\ref{app:appendix_A}, by means of 1000 mock catalogs, that cluster counts and clustering present negligible cross correlation for this kind of survey.

We thus consider the total likelihood as the product of three independent Gaussian likelihood functions, each taking the form
\begin{equation}    \label{eq:gauss_likelihood}
    \mathcal{L}(\mathbf{d} \,\vert\,\mathbf{m}(\boldsymbol{\theta}),\,C) = \frac{ \exp {\left \{-\frac{1}{2} [\mathbf{d} - \mathbf{m}(\boldsymbol{\theta})]^T C^{-1}  [\mathbf{d} - \mathbf{m}(\boldsymbol{\theta})] \right \}}}{\sqrt{(2\pi)^N \vert C \vert}} \,,
\end{equation}
where $\mathbf{d}$ if the data vector, $\mathbf{m}$ is the predicted quantity depending on cosmological and nuisance parameters $\boldsymbol{\theta}$, and $\mathbf{C}$ is the covariance matrix. Following \citet{Euclid:2021api} and \citet{Euclid:2022txd}, for cluster counts and clustering we assume the covariance to be cosmology dependent, computed through the analytical models described in the previous sections. For weak lensing log-masses, we assume a Gaussian likelihood with fixed-cosmology covariance, that accounts for shared multiplicative shear and photo-$z$ biases, blended sources, and cluster triaxiality and projection effects.

\begin{table*}
\centering           
\begin{tabular}{l c c c}
\hline
Parameter & Description & Prior  \\
\hline
$\om$        & Mean matter density   &  $[0.05,0.6]$ \vspace{0.1cm} \\ 
$\lnas$  & Amplitude of the primordial curvature perturbations   &  $[0.0,7.0]$ \vspace{0.1cm} \\ 
$\sigma_8$ & Amplitude of the matter power spectrum   &  -- \vspace{0.1cm} \\ 
$\logten M_{\rm min} \ [M_\odot/h]$ & Minimum halo mass to form a central galaxy  &  $[10.0, 14.0]$ \vspace{0.1cm} \\
$\logten M_1 \ [M_\odot/h]$  & Characteristic halo mass to acquire one satellite galaxy   &  $[10 M_{\rm min}, 30 M_{\rm min}]$ \vspace{0.1cm} \\ 
$\alpha$   & Power-law index of the richness–mass relation   &  $[0.1,1.5]$ \vspace{0.1cm} \\ 
$\sigma_{\rm intr}$ & Intrinsic scatter of the richness–mass relation   &  $[0.1,0.5]$ \vspace{0.1cm} \\ 
$s$       & Slope correction to the halo mass function   &  $\mathcal{N}(0.037, 0.014)$ \vspace{0.1cm} \\ 
$q$       & Amplitude correction to the halo mass function   &  $\mathcal{N}(1.008, 0.0019)$ \vspace{0.1cm} \\ 
$h$       & Hubble rate   &  $\mathcal{N}(0.7,0.1)$ \vspace{0.1cm} \\ 
$\Omega_{\rm b}\,h^2$    & Baryon density   &  $\mathcal{N}(0.02208, 0.0005)$ \vspace{0.1cm} \\ 
$\Omega_\nu\,h^2$  &  Energy density in massive neutrinos  &  $[0.0006, 0.01]$ \vspace{0.1cm} \\ 
$n_s$     & spectral index   &  $[0.8,1.2]$\\
\hline
\end{tabular}
\caption{\label{tab:sdss_prior} Model parameters and priors adopted in this analysis analysis. Parameters without a prior range are derived parameters. Ranges represent uniform flat priors, while $\mathcal{N}(\mu,\sigma)$ stands for Gaussian priors. } 
\end{table*}

\section{Data}\label{sec:data}
The dataset on which our analysis is based, as described in \citetalias{DES:2018crd}, is composed by 6964 photometrically selected galaxy clusters identified in the SDSS DR8, covering approximately 10\,000\,$\deg^2$. The cluster identification process relies on the redMaPPer cluster finding algorithm \citep{SDSS:2013jmz}, which models the red-sequence of galaxies and uses probabilistic richness estimation to identify clusters. The algorithm iteratively refines its model and employs percolation to connect galaxies into clusters. It estimates the purity and completeness of identified clusters and produces a catalog with cluster properties. The photometric redshift uncertainties for clusters are typically $\sigma_z /(1 + z) \approx 0.01$, and the analysis is confined to the redshift range $z \in [0.1, 0.3]$ to ensure the accuracy of our photometric measurements and maintain a well-defined volume-limited catalog. Only clusters of richness $\lambda \ge 20$ are considered, ensuring that 99 per cent of
the redMaPPer galaxy clusters are unambiguously mapped to
individual dark matter halos.

The observed number counts and weak lensing mass measurements are described in \citetalias{DES:2018crd}. Number counts are measured in five richness bins $\lob = \{20, 27.9, 37.6, 50.3, 69.3, 140\}$ and a single redshift bin $\zob \in [0.1, 0.3]$. Weak lensing masses are measured within the same intervals, from the shear catalog presented in \citet{Reyes2012a}, which includes $\sim$ 39 million galaxies over $\sim 9000\,\deg^2$ of the SDSS footprint. The mass estimates, obtained by fitting a NFW profile~\citep*{Navarro:1996gj} to the stacked lensing profile of clusters in the radial range from $0.3 \,h^{-1}$\,Mpc to $3 \,h^{-1}$\,Mpc, are a slightly updated from those presented in \citet{Simet:2016mzg}; the measurements account for a broad range of systematic uncertainties, including shear calibration and photo-z biases, dilution by member galaxies, source obscuration, magnification bias,
incorrect assumptions about the cluster mass profile, cluster miscentring, halo triaxiality and projection effect. 
Since the recovered weak lensing mass may not be identical to the mean mass of the clusters predicted by Eq.~\eqref{eq:mean_mass_wl}, due to the non-linear relation between the stacked density profile $\Delta \Sigma$ and the mass $M$, the relation between the recovered weak lensing mass and the mean mass in a bin has been calibrated through the use of simulations and applied directly to the observed data vector (see \citetalias{DES:2018crd} for details).

For a flat $\Lambda$CDM cosmology, Following \citetalias{DES:2018crd}, the cosmological dependence on $\om$ of the weak lensing mass estimates can be approximated by the linear relation in log-space:
\begin{equation} \label{eq:wl_mass}
    \logten \hat{M}_{\rm WL}(\om) = \logten \hat{M}_{\rm WL} \big \vert_{\om=0.3} + \frac{\diff \logten M_{\rm WL}}{\diff \om} (\om-0.3)\,,
\end{equation}
where the slopes derived in each bin from fitting such an equation to the data are listed in Table 1 of \citetalias{DES:2018crd}, and are used in our cosmological analysis to re-scale $\logten \hat{M}_{\rm WL}$ at each step of the MCMC. 

To measure the real-space 2-point correlation function, we consider 30 log-spaced separation bins in the range $r=20\,$--\,130\,$h^{-1}$\,Mpc; such an interval includes linear scales, where the bias is almost constant \citep{Manera:2009ak}, plus the BAO peak, whose position and amplitude are sensitive to the density parameters and $h$. Given the relatively low available statistics, we only consider a single richness threshold $\lob_{\rm th} \geq 20$, instead of binning over richness. We perform the measurement by applying the \citet{Landy:1993yu} estimator
\begin{equation}
    \hat{\xi}_{\rm h}^{aj} = \frac{{\rm DD}_{aj} - 2{\rm DR}_{aj} + {\rm RR}_{aj}}{{\rm RR}_{aj}}\,,
\end{equation}
where ${\rm DD}_{aj},\,{\rm DR}_{aj},\,{\rm RR}_{aj}$ are the number of pairs in the data-data, data-random and random-random catalogs within the $a$-th separation bin and $j$-th redshift bin, normalized for the number of objects in the data and random catalogs, $N_{\rm R}$ and $N_{\rm D}$ \citep{Kerscher:1999hc}. The measurement process is performed with the \texttt{CosmoBolognaLib} package \citep{Marulli:2015jil}. Differently from number counts and weak lensing masses, which are extracted from the ``v5.10'' of the SDSS redMaPPer cluster catalog, we measure cluster clustering from the ``v6.3'' catalog, for which the cluster random catalog is available. It has been verified that it has no significant impact on the results. The random catalog is generated by requiring that $f_{\rm mask} < 0.2$ and $\lambda / S > 20$, where $f_{\rm mask}$ is the local mask fraction and $S$ is a scale factor, as defined in \citet{SDSS:2013jmz}. The random catalog is 120 denser than the data catalog.  

\section{Results}\label{sec:results}
Assuming a flat $\Lambda$CDM cosmological model with massive (degenerate) neutrinos, we constrain six cosmological parameters $\om$, $\lnas$, $n_s$, $h$, $\Omega_{\rm b} h^2$, and $\Omega_\nu h^2$, plus the four mass-observable relation parameters of Eqs.~\eqref{eq:mor_mean} and~\eqref{eq:mor_var}, namely $\logten M_{\rm min}$, $\logten M_1$, $\alpha$, and $\sigma_{\rm intr}$. In addition, following \citetalias{DES:2018crd}, we add two parameters $s$ and $q$ to characterize the systematic uncertainty in the halo mass function fitted from simulations, such that
\begin{equation}
    \frac{\diff n}{\diff M} = \frac{\diff n}{\diff M}\,(s \logten(M/M^*) + q)\,,
\end{equation}
where $\logten(M^*) = 13.8\,h^{-1} M_\odot$, and $s=1$, $q=0$ are the reference values.
We assume the \citet{Euclid:2022dbc} model for the halo mass function (see Appendix~\ref{app:appendix_B} for details), and the \citet{Tinker:2010my} model for the halo bias.

The priors adopted for the parameter inference are listed in Table~\ref{tab:sdss_prior}, and the sampler employed for the analysis is the python wrapper for the nested sampler \texttt{PolyChord} \citep{Handley:2015fda}.

\subsection{Cosmological constraints} \label{sec:constraints}

\begin{figure*} 
    \centering
    \includegraphics[width=0.95\textwidth]{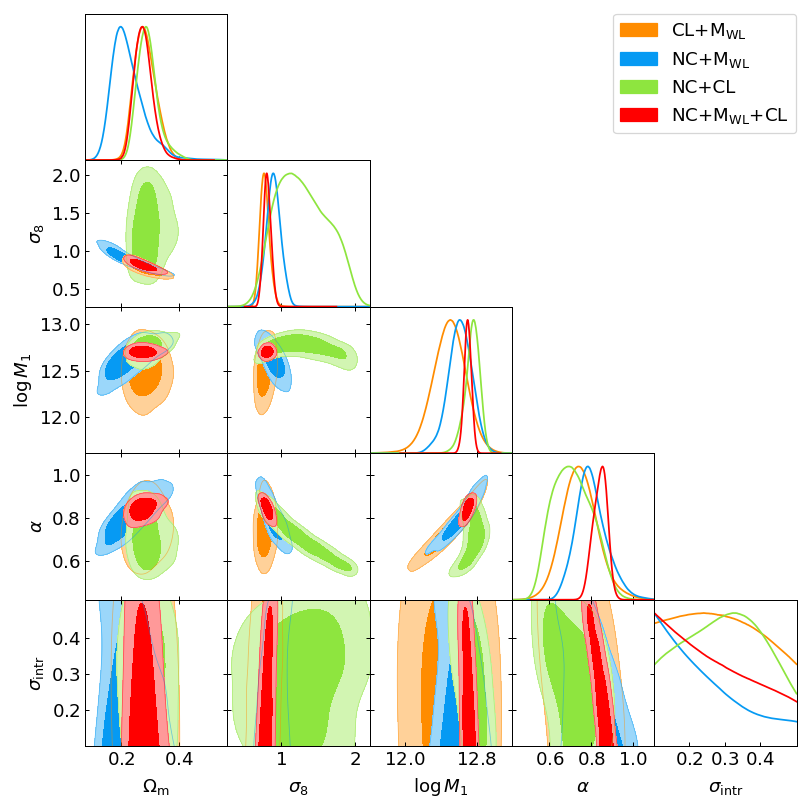}
    \caption{Contour plots at 68 and 95 per cent of confidence level for different observable combinations: number counts and weak lensing masses in blue, number counts and clustering in green, clustering and weak lensing masses in orange, and all the three probes in red. The posteriors are obtained marginalizing over the other parameters listed in Table~\ref{tab:sdss_prior}; $\sigma_8$ is a derived parameter.}
    \label{fig:prob_comp}
\end{figure*}

\begin{table*} 
\centering           
\begin{tabular}{l c c c c c c c}
\hline
Case & $\om$ & $\sigma_8$ & $\logten M_{\rm min}$ & $\logten M_1$ & $\alpha$ & $\sigma_{\rm intr}$ & $h$ \\
\hline
NC+$\mwl$    &  $0.22^{+0.03}_{-0.06} $  &  $0.90 \pm 0.10$  &  $11.28^{+0.14}_{-0.19}$  &  $12.61\pm 0.13$  &  $0.80^{+0.06}_{-0.08}$ &  $0.24^{+0.05}_{-0.14}$ & $0.72 \pm 0.08$ \vspace{0.2cm} \\ 
NC+CL        &  $0.29^{+0.03}_{-0.04}$  &  $1.30^{+0.29}_{-0.44}$  & $11.49^{+0.17}_{-0.19}$  &  $12.75^{+0.09}_{-0.06}$  &  $0.71^{+0.08}_{-0.11} $ &   $0.30\pm 0.11 $ & $0.63^{+0.03}_{-0.04}$ \vspace{0.2cm} \\ 
CL+$\mwl$    &  $0.28^{+0.03}_{-0.04}$ & $0.78^{+0.06}_{-0.07}$  &  $11.21\pm 0.23$  &  $12.49\pm 0.18$  &  $0.75^{+0.08}_{-0.09}$ &   $0.29^{+0.10}_{-0.15}$ & $0.63^{+0.03}_{-0.05}$ \vspace{0.2cm} \\ 
NC+CL+$\mwl$ &  $0.28 \pm 0.03$  &  $0.82 \pm 0.05$  &  $11.44^{+0.13}_{-0.16}$  &  $12.70\pm 0.04$  &  $0.84^{+0.04}_{-0.03}$  & $0.27^{+0.11}_{-0.16}$ &$0.64^{+0.03}_{-0.04}$\\
\hline 
\end{tabular}
\caption{\label{tab:sdss_bestfit} Best-fit values with 1$\sigma$ uncertainty for cosmological and mass-observable relation parameters for the four posteriors of Fig.\,\ref{fig:prob_comp}. We report here only parameters whose marginalized posteriors are not prior-dominated.}
\end{table*}

In the following we present cosmological constraints derived by carrying out four analyses, testing all the possible combinations of the three cosmological probes introduced in Sect. \ref{sec:theory}: 
\begin{enumerate}[label=\textit{\roman*)}]
    \item number counts and weak lensing masses (``NC+$\mwl$''), which represent the standard analysis performed by \citetalias{DES:2018crd};
    \item number counts and clustering (``NC+CL'');
    \item clustering and weak lensing masses (``CL+$\mwl$'');
    \item number counts, clustering and weak lensing masses (``NC+CL+$\mwl$'').
\end{enumerate}
In Appendix~\ref{app:appendix_B} we show the comparison of our NC+$\mwl$ analysis with the \citetalias{DES:2018crd} results, which demonstrates that the two analyses are fully consistent, plus some preliminary tests on the halo mass function and halo bias models.

In Table~\ref{tab:sdss_bestfit} we report the best-fit values with $1\sigma$ uncertainty obtained by the four analyses
, and in Fig.~\ref{fig:prob_comp} we compare the related posterior distributions on the cosmological parameters $\om$ and $\sigma_8$ (which is a derived parameter), and of the mass-observable relation parameters $M_1$, $\alpha$, $\sintr$; the results are obtained by marginalizing over all the other cosmological and mass-richness parameters.  From the figure, it can be noticed the different degeneracies between parameters constrained by different combinations of the three observables.  In particular, cluster clustering is effective in constraining the matter density parameter $\om$, breaking the $\om$\,--\,$\sigma_8$ degeneracy proper of cluster counts, and shifting its constraints towards higher values. The main reason for this is due to the sensitivity of the power spectrum to $\om$, which enters in determining its shape. More specifically, such information is mainly extracted from the BAO scales: we have verified, as shown in Appendix~\ref{app:appendix_C}, that restricting the clustering analysis to a separation range of $r \in [20 - 60]\,h^{-1}$\,Mpc significantly broaden the final posteriors, while considering the range $r \in [60 - 130]\,h^{-1}$\,Mpc allows us to recover almost all the information. Similarly, no additional information is gained by extending the clustering analysis to $r = 200\,h^{-1}\,$Mpc. 

Compared to other data combinations, the joint  analysis ``NC+CL'' (\textit{green} contours) has the least cosmological constraining power: the lack of ability in constraining the slope of the mass-observable relation translates into almost uninformative posteriors on $\sigma_8$. Indeed, the amplitude of the posterior on this parameter mainly depends on the amplitude of the prior on $A_s$. On the other hand, cluster clustering turns out to be extremely effective in constraining cosmology when combined with weak lensing mass information (\textit{orange} contours). In particular, the  $\om$ and $\sigma_8$ contours shrink by 22\,\% and 35\,\% compared to the ``NC+$\mwl$'' analysis. Despite the posteriors in the scaling relation parameters are somewhat weaker compared to the ``NC+$\mwl$`` combination, the break of the $\sigma_8$\,--\,$\alpha$ degeneracy allows this data combination to provide the tightest constraints in the $\om$\,--\,$\sigma_8$ plane (from the joint analysis of two observables). It should be noted that part of this constraining power comes from the clustering covariance matrix, which depends on shot-noise and thus contains information on the integrated halo mass function \citep[see][and discussion in Sect.~\ref{sec:cosmo}]{Euclid:2022txd}.

As expected by the presence of different degeneracies and parameter dependencies, the combination of the three observables (\textit{red} contours) provides the tightest posteriors  in whole parameter space. 
We obtain $\om = 0.28 \pm 0.03$ and $\sigma_8 = 0.82 \pm 0.05$.
Differently from the ``CL+$\mwl$'' case, which shows similar cosmological constraints, the three-probe combination also provides improved constraints on scaling relation parameters. Unsurprisingly, given the stacked approach used to estimate the mean cluster masses in the bins, the only scaling relation parameter that remains almost unconstrained is that describing the intrinsic scatter.

\begin{figure} 
    \centering
    \includegraphics[width=0.45\textwidth]{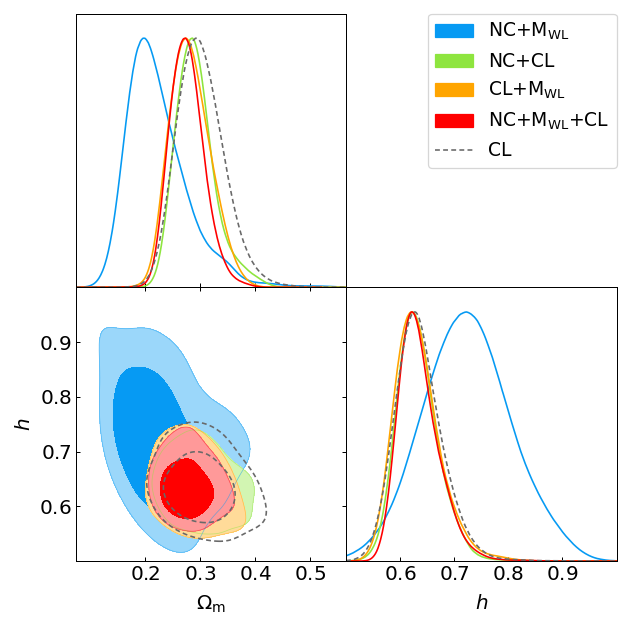}
    \caption{Contour plots at 68 and 95 per cent of confidence level in the $\om$\,--\,$h$ plane, for different observable combinations.}
    \label{fig:post_h}
\end{figure}

Besides breaking degeneracies when combined to the other statistics, cluster clustering also helps to constrain the Hubble parameter, as shown in Fig.~\ref{fig:post_h}. As indicated by the gray dashed contours, most of the information on the $h$ parameter is provided by the clustering alone, whose dependence on the expansion rate of the Universe is captured by the shape of the 2-point correlation function, in particular around the BAO scales \footnote{We assessed the influence of the adopted Gaussian prior on the $h$ posterior by repeating the analyses using a flat prior in the range $[0.4,1.0]$; the results do not change from using the Gaussian prior described in Table~\ref{tab:sdss_prior}.}. The best-fit value from the ``NC+$\mwl$+CL'' analysis is $h = 0.62^{+0.03}_{-0.02}$, with only a $1.90\sigma$ tension with the last Planck results \citep[$H_0 = 67.4 \pm 0.5\,{\rm km}\,s{\rm }^{-1}{\rm Mpc}^{-1}$, ][]{Planck:2018vyg}. Our constraints go in the opposite direction from that required to solve the tension with local probes, which prefer a higher value of $H_0$ \citep[e.g., $H_0 = (73.3 \pm 1.1)\,{\rm km}\,s{\rm }^{-1}{\rm Mpc}^{-1}$, ][]{Riess:2021jrx}.

On the other hand, we verified that the cluster clustering does not appear to carry additional information on the total neutrino mass, and even for the full data combination we retrieved posteriors on $\Omega_\nu$ consistent with the prior.

\subsection{Goodness of the fit and internal tensions}
\label{sec:goodfit}
\begin{figure*} 
    \centering
    \includegraphics[width=0.9\textwidth]{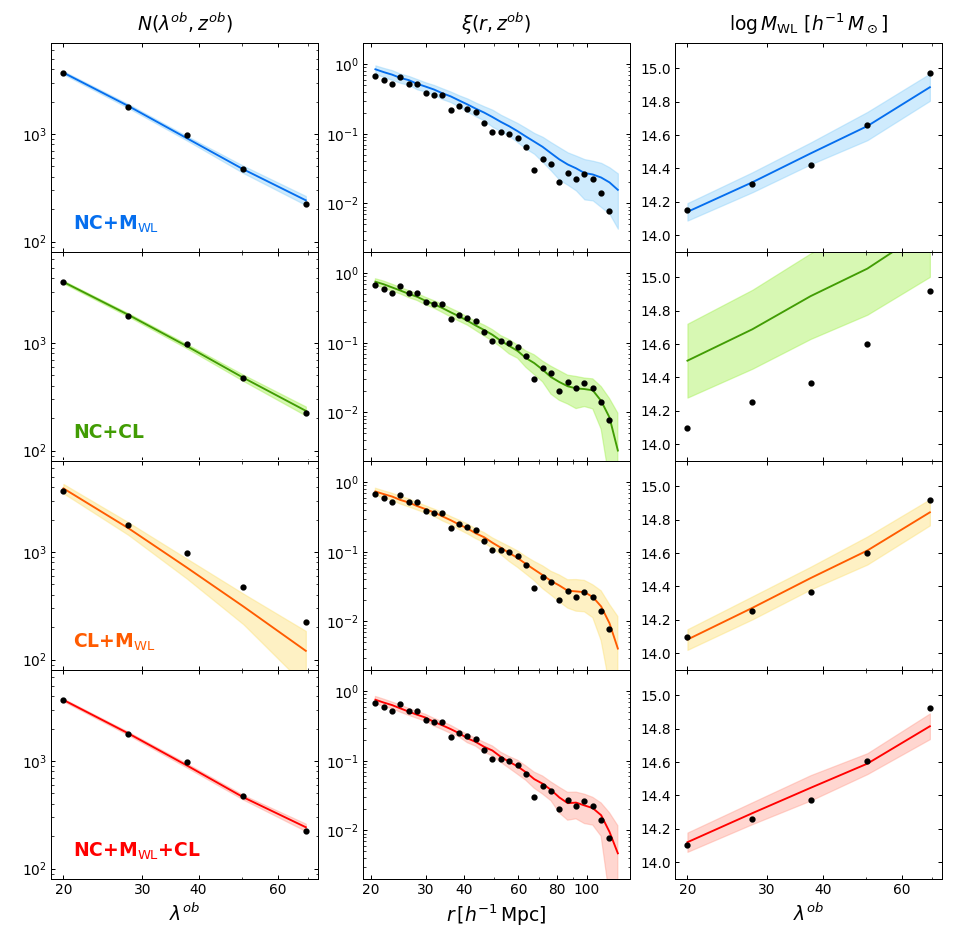}
    \caption{Observed number counts (\textit{left}), 2-point correlation function (\textit{middle}), and weak lensing masses (\textit{right}), compared to the corresponding predicted quantities evaluated at the best-fit cosmology of the contours in Fig.\,\ref{fig:prob_comp} (same color code). Black dost are the observed points, while colored lines and shaded areas represent, respectively, the mean and standard deviation of each statistics evaluated at 100 cosmologies randomly extracted from the posterior distribution, convoluted with the covariance matrix through a multivariate Gaussian distribution. }
    \label{fig:probes_bf}
\end{figure*}

In Fig.~\ref{fig:probes_bf}, we show with black dots the measured number counts (\textit{left} column), 2-point correlation function (\textit{middle} column) and mean cluster masses (\textit{right} column) compared to the corresponding predictions, computed by propagating the posteriors of Fig.~\ref{fig:prob_comp}, for the four different combination of probes: the solid lines and shaded areas represent the mean and standard deviation of each statistics evaluated over 100 random drawn from the corresponding posteriors; to also account for the uncertainty on the data, we convolve the predictions with the covariance matrix associated to the observable through a multivariate Gaussian distribution.  

For each data combination considered we obtain a good fit to data. In particular, the reduced $\chi^2$ for our best fit models are 1.91, 0.88, 0.94, and 1.04, for the ``NC+$\mwl$'', ``NC+CL'', ``CL+$\mwl$'', and ``NC+$\mwl$+CL'' joint analyses, respectively. In the cases where only two data sets are combined, we derive predictions for the third observable (first 3 rows of Fig.~\ref{fig:prob_comp}) to highlight possible internal tensions. For all these joint analyses, the predictions for the observable excluded from the fit are consistent within $2\sigma$ with the data. The worst agreement with the data is obtained for the mean cluster masses predicted by ``NC+CL'', and it is driven by the broad posterior on $\sigma_8$.

To quantitatively assess the level of tension among the data sets, we compute the posterior agreement \footnote{\url{https://github.com/SebastianBocquet/PosteriorAgreement}} \citep{SPT:2018njh}, which quantifies whether the difference between two posterior distributions is consistent with zero. We compute the agreement for the set of parameters \{$\om$, $\sigma_8$, $\alpha$, $M_1$\}, and we compare posteriors between pairs of probes.
The results are reported in Table~\ref{tab:sdpost_agreement}: the p-value indicates the probability of obtaining a difference between paired samples as extreme as, or more extreme than, the observed difference, under the assumption that there is no difference; a large (close to unity) p-value indicates a good agreement of the data sets.
The worst level of agreement is given by the comparison of ``NC+$\mwl$'' and ``NC+CL'', which can be interpreted as a small tension between clustering and weak-lensing masses: similarly, a tension of around 1$\sigma$ is obtained between number counts and weak-lensing masses (``NC+CL'' and ``$\mwl$+CL'' comparison), while a better agreement is reached by counts and clustering (``NC+$\mwl$'' and ``$\mwl$+CL'' comparison). However, none of the data combinations considered exhibit a statistically significant tension (i.e. $>3\sigma$).
This is also consistent with the rightmost panel in the second line of Fig.~\ref{fig:prob_comp}, which shows the largest tension between weak lensing masses and the prediction from ``NC+CL'' combination. While not statistically significant, these results might hint to some unmodeled systematics in the derivation of the mean weak lensing masses.

One systematic not accounted for in the analysis of \citet{Simet:2016mzg} is the optical selection bias induced by the correlation between richness and lensing signal \citep[see, e.g., ][]{DES:2020ahh, Sunayama:2020wvq, DES:2022qkn,Salcedo:2023aze}. We explicitly tested this hypothesis repeating the full joint analysis including a correction term in our mean mass model (Eqs.\ref{eq:mean_mass_wl}-\ref{eq:M_massfunc_obs}) which account for such lensing mass bias (see Appendix~\ref{app:appendix_D} for details). The analysis did not provide statistically significant evidences for the presence of biases in the mass estimates. On the other hand, this null result is in line with the fact that the full data combination does not show any internal tension.

\begin{table} 
\centering           
\begin{tabular}{l c c c} 
\hline 
Probes & p-value & agreement \\
\hline
NC+$\mwl$ - NC+CL     & 0.153 &  $1.43\,\sigma$ \vspace{0.1cm} \\ 
NC+$\mwl$ - CL+$\mwl$ & 0.822 &  $0.23\,\sigma$ \vspace{0.1cm} \\ 
$\mwl$+CL - NC+CL     & 0.407 &  $0.83\,\sigma$\\ 
\hline
\end{tabular}
\caption{\label{tab:sdpost_agreement} Posterior agreement between different probes combinations. The posterior agreement has been computed considering four parameters: $\om$, $\sigma_8$, $\alpha$, $M_1$.}
\end{table}


\subsection{Cosmology dependence of the covariance} \label{sec:cosmo}
In this section we discuss about the cosmology-dependence of covariance matrices. More specifically, the covariance models for number counts and clustering are assumed to be cosmology-dependent in the likelihood analysis, i.e., they are recomputed at each step of the MCMC process. This should represent the proper way to perform the analysis, as both the mean value and the covariance are needed to fully characterize the Gaussian model assumed to describe the data distribution \citep[e.g.][]{Krause:2016jvl,Eifler:2008gx,Morrison:2013tqa,Blot:2020cbi}. This correctly works for number counts, where the assumption of Gaussian likelihood is well motivated \citep[e.g.][]{Payerne:2022alz}. A Gaussian likelihood is also a common choice for the analysis of the two point correlation function, but there have been claims that the use of a cosmology-dependent covariance matrix may lead to an underestimation of the posteriors \citep{Carron:2012pw}. This approximation applied to the clustering of cluster statistic has been extensively tested in \citet{Euclid:2022txd}; the authors found that a Gaussian likelihood with a cosmology-dependent covariance matrix is statistically preferred over the fixed-cosmology case.

\begin{figure*} 
    \centering
    \includegraphics[width=0.99\textwidth]{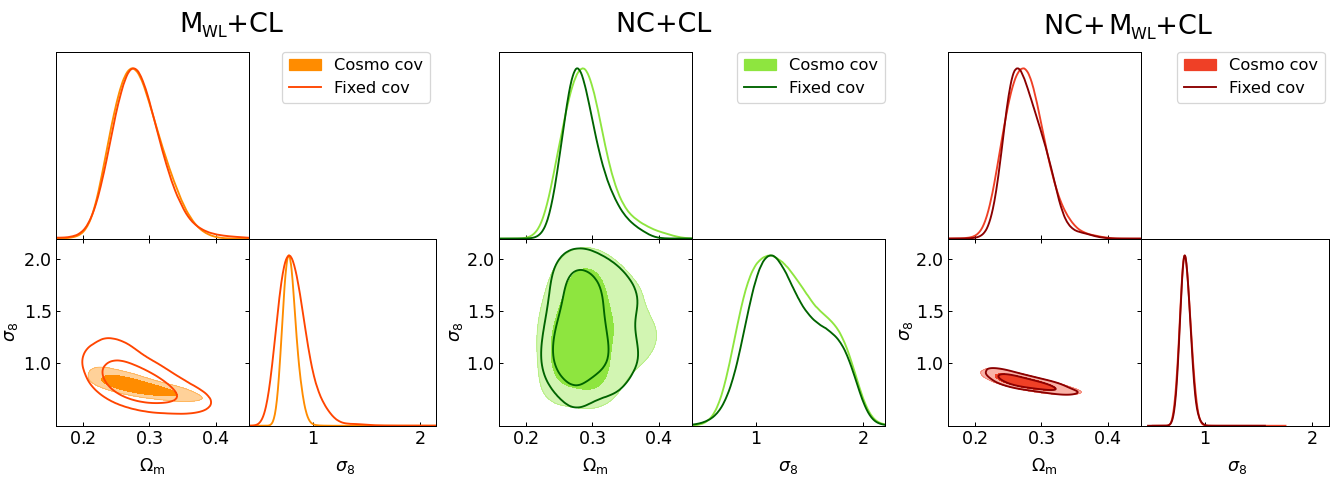}
    \caption{Comparison of $\om$\,--\,$\sigma_8$ posteriors obtained using a cosmology-dependent (filled contours) and fixed-cosmology (empty contours) clustering covariance. Different colors represent different probe combinations (color code as Fig.\,\ref{fig:prob_comp}). The fixed-cosmology covariance is computed at the best-fit parameters of the corresponding cosmology-dependent case.}
    \label{fig:prob_comp_cosmo}
\end{figure*}

To evaluate the effective impact of the cosmology-dependent clustering covariance on the SDSS redMaPPer cluster catalog used in our analysis, in Fig.~\ref{fig:prob_comp_cosmo} we compare the cosmological posteriors obtained by using a cosmology-dependent covariance matrix and by fixing it at a single cosmology. The latter has been chosen as the best-fit model obtained for the cosmology-dependent case. We perform this test for the three combinations of probes containing the cluster clustering. The most significant difference is observed for the ``CL+$\mwl$'' analysis, where the cosmology-dependent covariance matrix improves the constraints on $\sigma_8$ by approximately 100\,\%. As already mentioned in Sect.~\ref{sec:constraints}, the additional information comes from the shot-noise term proportional to the inverse of the mean density of halos, i.e., the integrated halo mass function. On the other hand, the posterior distributions of the other parameters remain mostly unaffected by the cosmology-dependent covariance matrix: this suggests that the integrated halo mass function provides additional information only on $\sigma_8$, while the dependence of the halo mass function on the other parameters is smoothed out by integration over the entire mass spectrum.

Conversely, when considering the combination ``NC+CL'',  the two posteriors look very similar. This is not surprising, since the cosmological information carried by the shot noise term through the halo mass function is already exhausted by the number count data, and thus the contribution of the cosmology-dependent covariance matrix becomes negligible. In this context, it could be argued that the information of the two observables is counted twice: : as already stated, the clustering covariance includes a shot-noise contribution containing part of the information of the number counts; viceversa, the number counts covariance contains part of the clustering information, through the sample variance term. However, such terms do not directly affect the mean values, but contribute to modulate the fluctuations around the mean, thus characterizing the covariance matrices of the two statistics. It is noteworthy that if there were a correlation that is erroneously overlooked, the halo mass function information would be counted twice and the cosmology-dependent contours of the ``NC+CL'' case (\textit{dark green} lines) should tighten with respect to to the fixed-cosmology contours (\textit{light green} contours), which is not the case. For the same reason, there is no difference between the cosmology-dependent and fixed-cosmology cases when combining all the three probes.

\section{Discussion and conclusions}\label{sec:conclusions}

\begin{figure} 
    \centering
    \includegraphics[width=0.51\textwidth]{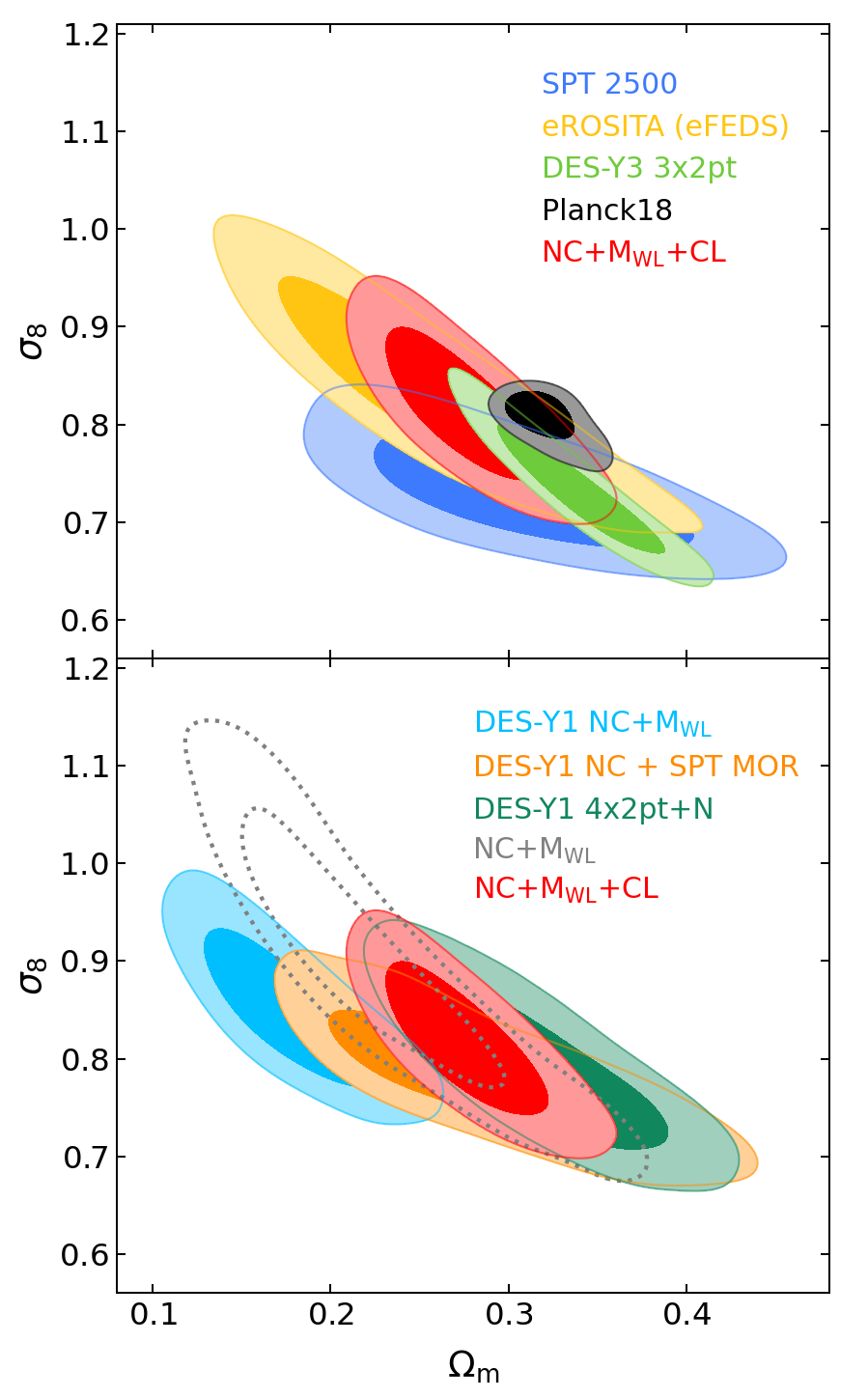}
    \caption{Comparison of $\om$\,--\,$\sigma_8$ posteriors obtained by different surveys. \textit{Top panel}: 2500\,$\deg^2$ SPT-SZ cluster survey \citep[][\textit{blue} contours]{SPT:2018njh}, eROSITA (eFEDS) cluster survey \citep[][\textit{yellow} contours]{Chiu:2022qgb}, DES-Y3 NC+3x2pt \citep[][\textit{green} contours]{DES:2021wwk}, and Planck TT,TE,EE+lowE with free $m_\nu$ \citep[][\textit{black} countours]{Planck:2018vyg}, compared to our results ``NC+$\mwl$+CL'' (\textit{red} contours) analyses. \textit{Bottom panel}: DES-Y1 number counts and weak lensing mass estimates \citep[][\textit{cyan} contours]{DES:2020ahh}, DES-Y1 cluster counts data and SPT-SZ follow-up data \citep[][\textit{orange} contours]{DES:2020cbm}, DES-Y1 NC+4x2pt \citep[][\textit{dark green} contours]{DES:2020mlx}, compared to our results from ``NC+$\mwl$'' (\textit{dotted gray} contours) and ``NC+$\mwl$+CL'' (\textit{red} contours) results.}
    \label{fig:survey_comp}
\end{figure}

\begin{figure} 
    \centering
    \includegraphics[width=0.51\textwidth]{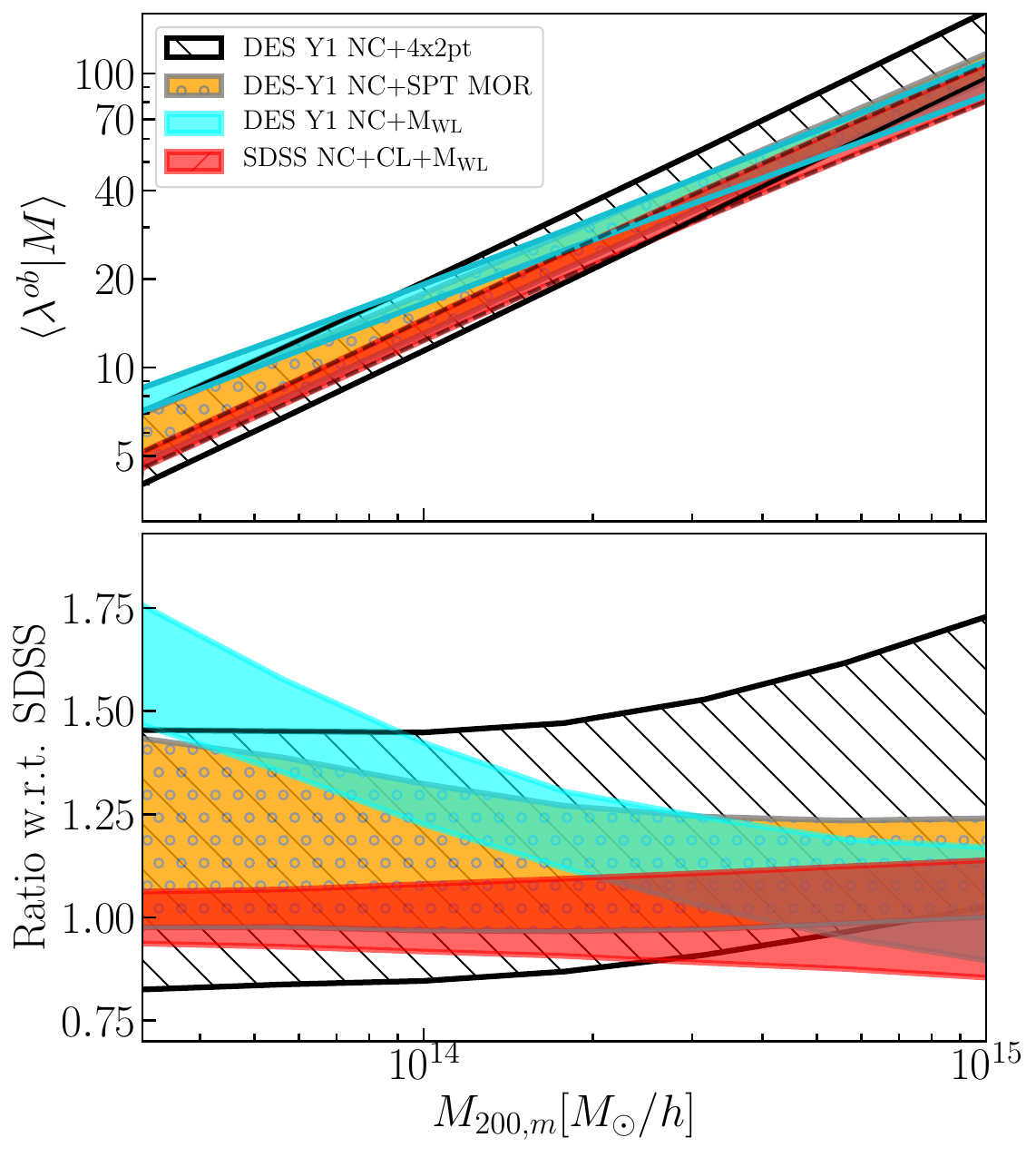}
    \caption{Comparison of the mean richness-mass relations derived in this work (\textit{red} band) and those derived from different analyses of the DES-Y1 redMaPPer cluster sample: DES-Y1 number counts and weak lensing mass estimates \citep[][\textit{cyan} band]{DES:2020ahh}, DES-Y1 cluster counts data and SPT-SZ follow-up data \citep[][\textit{orange-dotted} band]{DES:2020cbm}, and DES-Y1 cluster abundance with galaxy, lensing and cluster auto and cross-correlation function \citep[][\textit{black-hatched} band]{DES:2020mlx}. The bands width corresponds to $1\sigma$ uncertainty of the mean relation. \textit{Lower} panel: Ratio of the DES-Y1 scaling relations with respect to the SDSS result.}
    \label{fig:lob_comp}
\end{figure}

In this work we have extended the analysis described in \citetalias{DES:2018crd}, by adding to numbers counts and weak lensing mass estimates also the cosmological information contained in the 2-point correlation function of galaxy clusters. The twofold purpose of our analysis is to determine whether the inclusion of cluster clustering can help from one hand to improve cosmological constraints, and, on the other hand, to highlight possible tensions between the data sets. To this purpose, we analyzed the redMaPPer cluster catalog from the SDSS DR8 \citep{SDSS:2011gir}, containing 6964 clusters with richness $\lob \ge 20$ and redshift $\zob \in [0.1, 0.3]$. Assuming a flat $\Lambda$CDM model with massive neutrinos, we constrained cosmological and mass-observable relation parameters using different combinations of the probes: cluster number counts, cluster clustering, and weak lensing masses. 

The combination of all the three probes has proved to be very constraining for the parameters defining both the underlying cosmological model and the mass-richness relation. The inclusion of cluster clustering helps to better constrain $\om$ and shifts its posterior toward higher values. Moreover, it allows to obtain a precise measurement of the Hubble parameter ($h=0.64^{+0.03}_{-0.04}$), which is independent and competitive to other probes. Instead, cluster clustering does not prove useful in constraining neutrino mass any more than does the ``NC+$\mwl$'' combination; in fact, in all the cases the posterior on the neutrino mass are consistent with its prior.

Cluster clustering combined with number counts alone does not carry enough information on the scaling relation to compensate for the lack of dedicated mass-calibration data, and this comes at the expense of very weak constraints on $\sigma_8$. Instead, the different parameter degeneracies of clustering and weak lensing masses allows their combination to constrain cosmology better than the standard combination of number counts and weak lensing masses. It is interesting to notice that, while the amplitude of $\om$\,--\,$\sigma_8$ contours is comparable with the three-probe case, the scaling relation parameters are much less constrained when number counts are not included in the fit; this indicates that the combination of clustering and weak lensing masses without the full number counts information can tightly constrain cosmology even with a relatively limited knowledge of the scaling relation compared to the standard ``NC+$\mwl$'' analysis.
However, even if number counts are not included in the analysis, part of the halo mass function information is contained in the cosmology-dependent covariance matrix entering in the 2-point correlation function likelihood. Indeed, we showed that such information is useful to obtain better parameter constraints when number counts are not directly included in the inference process.

In Fig.~\ref{fig:survey_comp} we present a comparison of our contours with constraints obtained from other probes and from other cluster surveys. In the \textit{top panel}, we consider the outcomes of the following analyses: \citet{SPT:2018njh} analyzed the cluster count data from the 2500\,$\deg^2$ SPT-SZ catalog in combination with high-quality X-ray and lensing follow-up data for 121 systems; \citet{Chiu:2022qgb} examined cluster abundances from the X-ray sample identified in the eROSITA Final Equatorial Depth Survey (eFEDS) in combination with Hyper Supreme Camera weak lensing data; \citet{DES:2021wwk} combined cosmic shear, galaxy clustering and galaxy-galaxy lensing from the third year release of DES data (DES-Y3); and the baseline analysis of the CMB anisotropies with free neutrino mass conducted by \citet{Planck:2018vyg}. In the \textit{bottom panel} we compare our results of the first year release of the DES cluster sample data: \citet{DES:2020mlx} combined DES-Y1 cluster number counts and galaxy, shear and cluster auto and cross correlation functions; \citet{DES:2020cbm} combined the DES-Y1 cluster counts data with SPT-SZ multi-wavelength data; \citet{DES:2020ahh} combined DES-Y1 number counts and weak lensing mass estimates.
Our ``NC+$\mwl$+CL'' results in the $\sigma_8-\om$ plane are in agreement with all the other analyses considered within one sigma, but for the $\om$ value preferred by Planck and \citet{DES:2020ahh}, which are at 1.22\,$\sigma$ and 2.21\,$\sigma$ tension with our results. Also, our results show a small tension with the ones of \citet{DES:2021wwk}, corresponding to 1.35\,$\sigma$ on $\om$ and 1.30\,$\sigma$ on $\sigma_8$. The constraining power is competitive with those of other recent cluster abundance studies. The comparison with DES-Y1 NC+4x2pt results is particularly interesting, as the two cluster samples have similar sizes, and the observables considered in this analysis mirror those analyzed in the work of \cite{DES:2020mlx} (besides the galaxy clustering and galaxy-cluster cross-correlations). It is worth mentioning that our contours fully match the DES results (at 0.45$\sigma$ and 0.47$\sigma$ level for $\om$ and $\sigma_8$, respectively), but the amplitude is more constrained. A possible explanation for the larger contours found in \cite{DES:2020mlx} are the different scales considered in the 2 point correlation functions: the analysis 
is limited up to a scale of $100\,h^{-1}\,$Mpc, so that part of the information carried by the BAO peak is lost. In this work, the clustering is measured up to $130\,h^{-1}\,$Mpc, allowing the full exploitation of the BAO feature (see Appendix~\ref{app:appendix_C}). On the lower end, they do not use data below $8\,h^{-1}\,$Mpc, thus excluding the 1-halo term information from the lensing analysis and reducing significantly the capability of constraining  the cluster masses. Moreover, \cite{DES:2020mlx} considers a selection bias term, which modulates the amplitude of the  cluster auto and cross-correlation functions, further reducing the constraining power of the sample.

In Fig.~\ref{fig:lob_comp} we compare the mean richness-mass relation derived from our ``NC+$\mwl$+CL'' analysis, with those obtained from the analyses of the DES-Y1 cluster sample considered above. For the comparison, all the DES-Y1 relations have been evolved to the mean redshift of the SDSS sample, $z=0.22$, and re-scaled by the factor $0.93$ to correct for the systematic richness offset between SDSS and DES-Y1 catalogs \citep[see][]{McClintock2019}.
All the scaling relations show good agreement over the relevant mass range, $M\gtrsim 10^{13.5} h^{-1}M_\odot$, but the \citet{DES:2020ahh} results, which exhibits a shallower slope and deviate from the other relations by $1\sigma$ in the low-mass end. As discussed in \citet{DES:2020ahh} this tension is likely due to a flawed modeling of the stacked weak lensing signal, and it drives the tension with other cosmological probes. It is interesting to notice that despite the similar technique and modeling adopted to estimate the weak lensing masses in this work and in \citet{DES:2020ahh}, the SDSS result does not seem to be affected by the same systematic. A possible explanation is the different scale cut adopted in the lensing analyses: \citet{Simet:2016mzg} considers only scales below $R \simeq 3 h^{-1}$Mpc, while \citet{McClintock2019} includes scales up to $30 h^{-1}$Mpc modeling also the two-halo term. These large scales are the most affected by the optical selection bias discussed in Sect.~\ref{sec:goodfit} \citep[see e.g.][]{Sunayama:2020wvq,DES:2022qkn}, which represents one of the main systematic in \citet{DES:2020ahh}.
Conversely, \citet{DES:2020mlx} excluded the 1-halo term from the analysis by considering $R>8 h^{-1}$Mpc, but included a selection bias correction term left free to vary in the analysis, which can explain the good agreement with our result. Finally, the weak lensing follow-up data used in \citet{DES:2020cbm} to calibrate the scaling relations, being based on a SZ-selected cluster sample, is not affected by such selection bias.

Finally, it is worth mentioning the comparison with the work of \citet{Sunayama:2023hfm}, who also analyzed the SDSS redMaPPer cluster catalog in combination with the Hyper-Suprime Cam (HSC) Year3 shape catalog, by combining cluster abundances, weak-lensing masses, and projected cluster clustering.  We get consistent results, however, their posterior amplitude is slightly larger. Again, this is mainly ascribed to the different scales used for cluster clustering: in their analysis the clustering is measured up to $50\,h^{-1}\,$Mpc, thus excluding the BAO peak that carries most of the information. Indeed, the broadening of the contours  generated by examining the range of separation $r \in [20 - 60]\,h^{-1}\,$Mpc shown in Fig.~\ref{fig:scales_cl} (blue contours) is wide enough to be consistent with the amplitude of their posteriors. Moreover, they include a selection bias term which modulates both the lensing and the 2-point correlation function amplitude.

In summary, in the analysis presented here we have shown some of the potential of cluster clustering, which has proven to be useful for constraining cosmology, although the catalog analyzed still has rather limited statistics and a narrow redshift coverage. To better exploit the clustering information, a full redshift- and richness-dependent analysis should be performed. Also, the inclusion of redshift-space distortions in the modeling of the 2-point function may further increase the constraining power. Finally, exploring cluster clustering analysis within the context of non-standard cosmological models is an interesting avenue of research. This approach allows us to investigate the impact of phenomena like dark energy or modified gravity on the formation and evolution of the large-scale structure of the Universe. In fact, these non-standard models have the potential to induce substantial alterations in the clustering patterns of cosmic structures, including that of galaxy clusters. Notably, in such non-standard scenarios, the halo bias may acquire a scale-dependent character, offering a valuable way for probing and constraining the underlying cosmological framework through cluster clustering analysis.

We expect that future large survey catalogs, such as \textit{Euclid} or LSST, will make it possible to measure cluster clustering with growing accuracy, making its combined analysis with other cosmological probes increasingly powerful and competitive.

\section*{Acknowledgements}
AS, TC and SB are supported by the INFN INDARK PD51 grant. AF acknowledges support from Brookhaven National Laboratory. AS is also supported by the ERC `ClustersXCosmo' grant agreement 716762. TC and AS are also supported by the FARE MIUR grant `ClustersXEuclid' R165SBKTMA. 

\bibliographystyle{mnras}
\bibliography{biblio}

\appendix
\section{Cluster counts and clustering cross-correlation matrix} \label{app:appendix_A}
In this appendix we discuss about the cross-correlation between number counts and clustering. In principle such observables present a correlation due to density fluctuations on scales larger than the survey size, i.e. the super-sample covariance \citep[see, e.g., ][]{Takada:2013wfa, Krause:2016jvl}. The effect of super-sample covariance only depends on the survey size, and should become negligible for large volumes. Moreover, it mostly impact small, non-linear scales. 

\begin{figure} 
    \centering
    \includegraphics[width=0.5\textwidth]{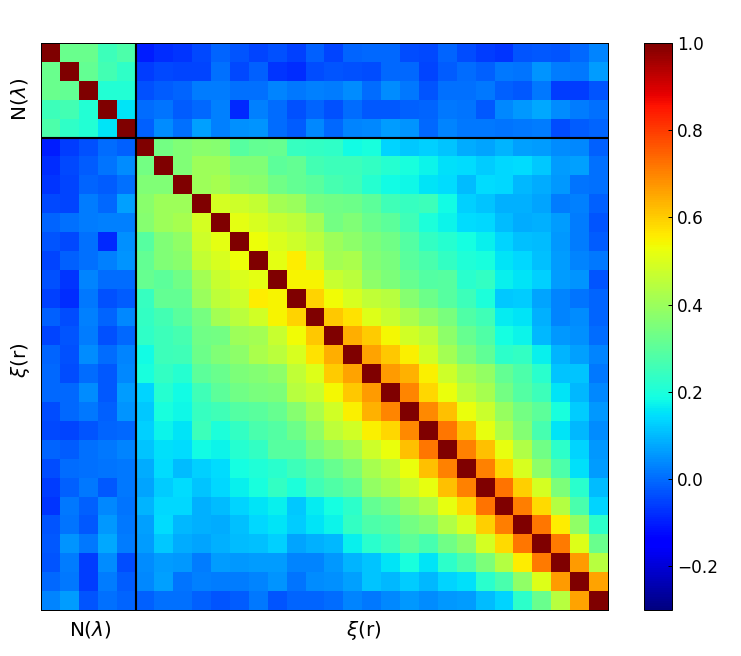}
    \caption{Normalized covariance matrix between cluster counts and cluster clustering. The block-diagonal terms represent the auto-correlation matrix for number counts (\textit{upper left}) and clustering (\textit{lower right}), while the off-diagonal blocks are the cross-correlation matrix.}
    \label{fig:covariance}
\end{figure}

To assess the entity of such a correlation on a survey with properties similar to those analyzed in this work, we make use of a set of 1000 lightcones simulated with the PINOCCHIO code \citep{Monaco:2001jf,Munari:2016aut}, that are described in detail in \citet{Euclid:2021api}. The lightcones cover an area of $\sim 13\;000\,\deg^2$, which is comparable with the area of the cluster catalog analyzed in this work, and we consider the redshift range $z \in [0.1, 0.3]$. We measure number counts and 2-point correlation function from each one of the 1000 catalogs, and use them to compute a numerical covariance matrix, whose normalized version is shown in Fig.~\ref{fig:covariance}. The block-diagonal elements represent the cluster counts (\textit{upper left}) and cluster clustering (\textit{lower right}) covariances, while the off-diagonal elements are the cross-covariance between the two statistics. We can notice that the cross-correlation is entirely dominated by noise, and consistent with zero signal. 

This result suggests that the two observables can be considered ad independent. A similar outcome has been obtained in \citet{Sunayama:2023hfm}.

\section{Validation tests}\label{app:appendix_B}
In this appendix we present the results from tests aimed at assessing the validation of our analyses. 

\begin{figure*} 
    \centering
    \includegraphics[width=0.62\textwidth]{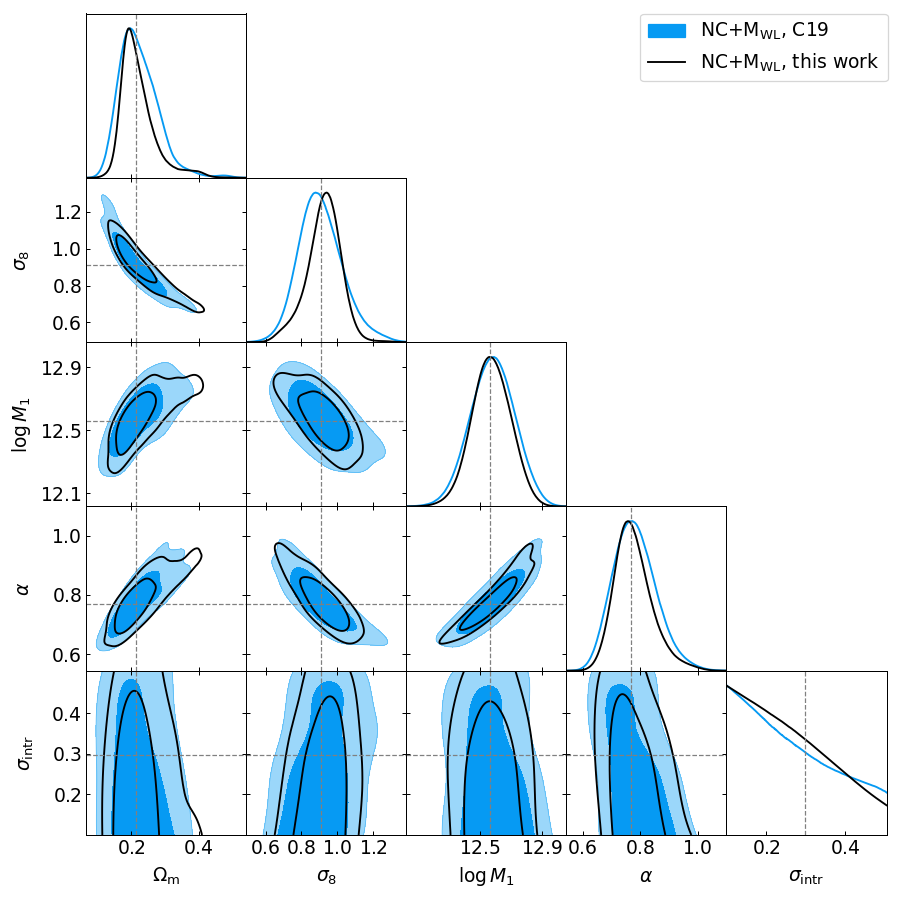}
    \caption{Comparison between \citetalias{DES:2018crd} results and this work. Gray dashed lines are the best-fit parameters from \citetalias{DES:2018crd}.}
    \label{fig:DES19_comp}
    \bigskip 
    \includegraphics[width=0.62\textwidth]{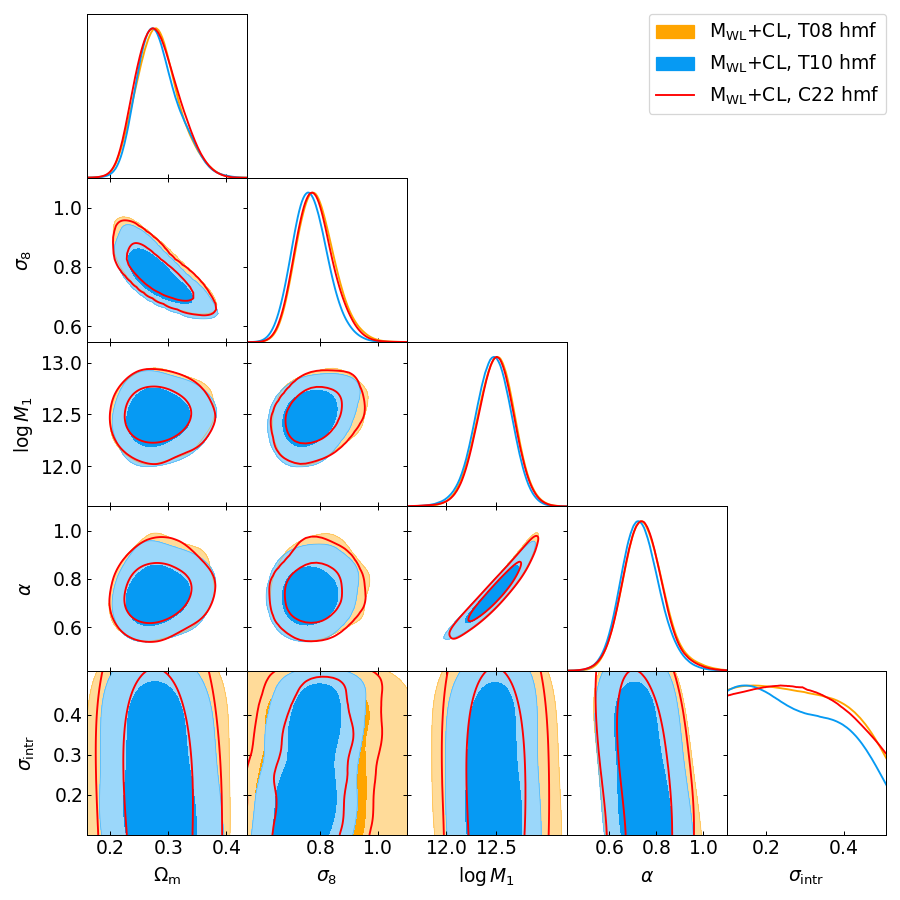}
    \caption{Comparison between different halo mass function models from the ``$\mwl$+CL'' analysis: \citet{Tinker:2008ff} in orange, \citet{Tinker:2010my} in blue, and \citet{Euclid:2022dbc} in red.}
    \label{fig:hmf_comp}
\end{figure*}

In Fig.~\ref{fig:DES19_comp} we compare our results for the ``NC+$\mwl$'' combination with the same posterior distribution from \citetalias{DES:2018crd}. 
By comparing our results (solid black contours) with \citetalias{DES:2018crd} results (blue shaded area), we notice a general good agreement, with only a small difference in the posteriors' amplitude. This difference can be ascribed to numerical issues due to the use of a different code. Indeed, although the likelihood and sampler adopted are the same, it is unlikely that two different codes will achieve the same numerical accuracy, and this may produce sizable differences in the predictions of the observables and, consequently, in the cosmological constraints. However, we reiterate that the two results from our and \citetalias{DES:2018crd} analyses are fully consistent with each other. This consistency check ensures the absence of additional systematics in our analysis with respect to the \citetalias{DES:2018crd} case, confirming a good agreement between the two analyses.

We also assess the impact of different halo mass function models, and of their interaction with the halo bias model. In fact, the halo bias is derived from the halo mass function through the peak-background split formalism \citep{Cole:1989vx,Mo:1995cs}, and is thus related to the mass function model from which is obtained. In this work, we test three parametrizations for the halo mass function, i.e., \citet{Tinker:2008ff}, \citet{Tinker:2010my}, and \citet{Euclid:2022dbc}, while we only assume a single model for the halo bias, from \citet{Tinker:2010my}.

In Fig.~\ref{fig:hmf_comp} we show the posteriors resulting from the use of different halo mass function parametrizations: orange contours refer to the \citet{Tinker:2008ff} model, blue contours to the \citet{Tinker:2010my} model, and red contours to the \citet{Euclid:2022dbc} model. We show here the constraints from the ``$\mwl$+CL'' analysis, to better estimate the impact of using the same halo bias in combination with different halo mass function models; the results from the other combinations of observables are consistent with those presented here. By comparing the results, we notice that the halo mass function model does not have any significant impact on parameter constraints at this level of statistics. Neither the use of a bias model not associated with the halo mass function produces measurable effects on the contours. From this result we conclude that the choice of the halo mass function can be made without any particular constraint. Therefore, in this work we choose to use the \citet{Euclid:2022dbc} model, which has been demonstrated to be accurate at the sub-percent level when tested on  numerical simulations.

\section{Clustering scales} \label{app:appendix_C}
\begin{figure*} 
    \centering
    \includegraphics[width=0.7\textwidth]{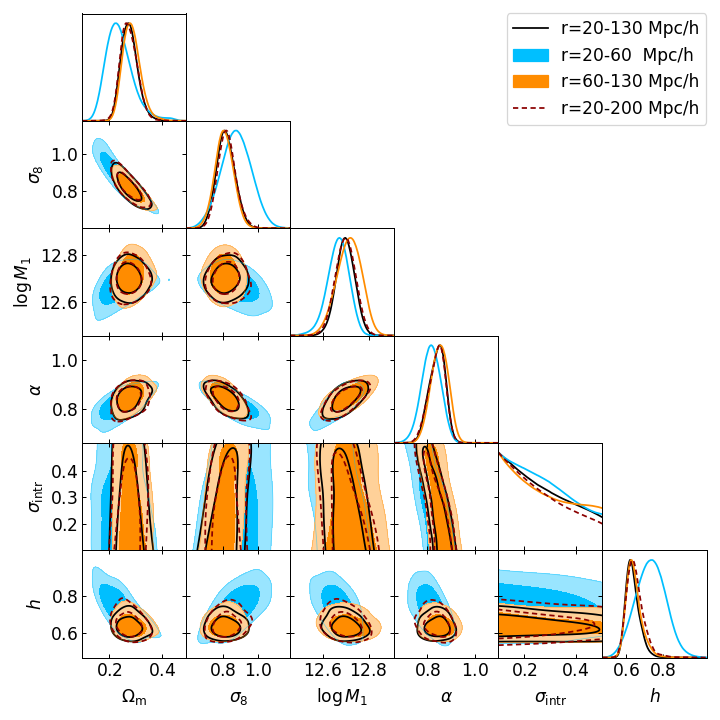}
    \caption{Contour plots at 68 and 95 per cent of confidence level, corresponding to different separation ranges in the cluster clustering modeling. In all the cases, the contours correspond to the ``NC+$\mwl$+CL'' combination.}
    \label{fig:scales_cl}
\end{figure*}

We show in Fig.~\ref{fig:scales_cl}, the cosmological constraints from the ``NC+$\mwl$+CL'' analysis, obtained by varying the clustering scales. The standard separation range adopted through this paper is given by $r \in [20 - 130]\,h^{-1}$\,Mpc (black empty contours). We split such a range in order to quantify the impact of the BAO scales on the final parameter constraints; in such a way, we obtain a low-scales interval $r \in [20 - 60]\,h^{-1}$\,Mpc (blue filled contours), and high-scales interval $r \in [60 - 130]\,h^{-1}$\,Mpc (orange filled contours). The comparison highlights that the low-scale part of the (linear) 2PCF has much less predictive power than that measured around the BAO peak, which actually carries most of the information. A similar outcome is found by expanding the range beyond the BAO scales, i.e., $r \in [20 - 200]\,h^{-1}$\,Mpc: the contours remain almost unchanged, indicating that no further information can be extracted from the 3D cluster clustering. 
The latter test also assures us that we are not affected by border effects when modeling the 2PCF, as no systematics due to the assumption of spherical shells (see Eq.~\ref{eq:shell_window}) are added when increasing the separation up to $200\,h^{-1}$\,Mpc.

\section{Richness and weak lensing correlation}\label{app:appendix_D}
We attempted to get leverage of the clustering information and constrain a possible mass bias induced by the correlation of the richness and lensing signal (see Sect.~\ref{sec:goodfit}). In particular we assume the mean weak lensing mass and richness to follow a bivariate log-normal distribution. With this assumption we can correct the mean mass predictions adding to Eq.~\eqref{eq:M_massfunc_obs} the bias term:
\begin{equation}
    \rho \, \sigma_{\mwl} \, \frac{\ln \ltr - \langle \ln \ltr | M \rangle}{\sigma_{\ln \ltr}}\,,
\end{equation}
where $\rho$ is the correlation coefficient between richness and weak lensing mass, $\sigma_{\mwl}$ is the intrinsic scatter at fixed mass of the latter, , and $\langle \ln \ltr | M \rangle$ and $\sigma_{\ln \ltr}$ are given by Eqs.~\eqref{eq:mor_mean} and \eqref{eq:mor_var}. Note that for this exercise, besides assuming a log-normal distribution for the richness-mass relation $P(\ltr | M, \ztr)$ we neglect the observational scatter in the richness mass relation, $P(\lob | \ltr, \ztr)$ in Eq.~\eqref{eq:mor}, assuming instead $\lob = \ltr$; we verified that these two approximations bring a negligible variation of the posteriors. In addition to the parameters listed in Table~\ref{tab:sdss_prior}, we set $\rho$ and $\sigma_{\mwl}$ to be free parameters with priors given by $\rho = [-1, +1]$ and $\sigma_{\mwl} = \mathcal{N}(0.2,0.1)$.  The three probe combination ``NC+CL+$\mwl$'' was examined. The analysis is not conclusive as the two parameters remain basically unconstrained. The lack of meaningful constraints derived on these two parameters is in line with the result that tensions disappear in the analysis based on the combination of the three probes.
\end{document}